\documentclass[11pt,a4paper]{article}
\usepackage[english]{babel}
\usepackage{a4wide}
\usepackage[latin1]{inputenc}
\usepackage{amsmath}
\usepackage{amsfonts}
\usepackage{amssymb}
\usepackage{epsfig}
\usepackage{graphicx}
\usepackage{latexsym}
\usepackage{color}

\title{Goal Clustering: VNS based heuristics}
\author{Pedro Martins \\ Center for Mathematics, Fundamental Applications and \\ Operations Research (CMAF-CIO) - University of Lisbon, Portugal
and \\ Polytechnic Institute of Coimbra - ISCAC, Portugal \\
e-mail: \texttt{pmartins@iscac.pt}}

\newtheorem{prop}{Proposition}

\newtheorem{cor}{Corollary}

\begin{document}

\maketitle \setcounter{page}{1}

\begin{abstract}
Given a set $V$ of $n$ elements on $m$ attributes, we want to find a partition of $V$ on the minimum number of clusters such that the associated \emph{R-squared} ratio is at least a given threshold. We denote this problem as \emph{Goal Clustering} (GC).

This problem represents a new perspective, characterizing a different methodology within unsupervised non-hierarchical clustering. In effect, while in the \emph{k}-means we set the number of clusters in advance and then test the associated \emph{R-squared} ratio; in the GC we set an \emph{R-squared} threshold lower limit in advance and minimize $k$.

We present two Variable Neighborhood Search (VNS) based heuristics for the GC problem. The two heuristics use different methodologies to start the VNS algorithms. One is based on the Ward's construction and the other one resorts to the \emph{k}-means method.

Computational tests are conducted over a set of large sized instances in order to show the performance of the two proposed heuristics.

\medskip
\small \noindent \textbf{Keywords:} non-hierarchical clustering, unsupervised clustering, \emph{k}-means, Ward's algorithm, \emph{R-squared} ratio, VNS metaheuristics

\medskip
\small \noindent \textbf{Mathematics Subject Classification:} 62H30, 90C59, 90C27, 05C70
\normalsize

\end{abstract}

\section{Introduction} \label{sec:Intro}
Given a set $V$ of $n$ points/elements on $m$ attributes, with $x_{ij}$ representing the $j^{\scriptsize\mbox{th}}$ attribute of point $i$, for $i=1,...,n$ and $j=1,...,m$; the \emph{Goal Clustering} (GC) problem is to find a partition of the set of points into cohesive groups/clusters/components such that the associated \emph{R-squared} ratio is at least a given threshold. The goal of the problem is to find a partition with the minimum number of groups, satisfying the mentioned \emph{R-squared} lower limit threshold. In what follows, we denote the \emph{R-squared} ratio by $R^2$ and the threshold by $R^2T$.

A partition of $V$ on $k$ groups is characterized by $V=Q^1 \cup Q^2 \cup \ldots  \cup Q^k$, with $Q^a \cap Q^b = \emptyset$ for all $a,b = 1,...,k$ and $Q^a \neq \emptyset$ for all $a=1,...,k$.

For a given partition $V=Q^1 \cup Q^2 \cup \ldots  \cup Q^k$, the $R^2$ ratio represents the proportion of the total variability retained between the groups, being defined by
\begin{equation}\label{eqn:c1}
    \displaystyle R^2 = \frac{SSB}{SST}
\end{equation}
with $SSB = \sum_{j=1}^m SSB_j$ and $SST = \sum_{j=1}^m SST_j$. Function $SSB_j$ represents attribute's $j$ between-groups variability and $SST_j$ is attribute's $j$ total variability, with
\begin{equation}\label{eqn:c2}
    SSB_j = \displaystyle\sum_{q=1}^k |Q^q| \cdot \left(\overline{x}_q^j-\overline{x}^j \right)^2\;, \mbox{  for all } j=1, \ldots m
\end{equation}
and
\begin{equation}\label{eqn:c3}
    SST_j = \displaystyle\sum_{i \in V} \left(x_{ij}-\overline{x}^j \right)^2\;, \mbox{  for all } j=1, \ldots m
\end{equation}
$\overline{x}_q^j$ represents the average value for attribute $j$ among all points in group $Q^q$; and $\overline{x}^j$ represents the average value for attribute $j$ among all points in $V$.

The $R^2$ ratio can also be represented in terms of the within-groups variability $SSW$, as
\begin{equation}\label{eqn:c4}
    \displaystyle R^2 = 1 - \frac{SSW}{SST}
\end{equation}
given that $SST_j = SSW_j + SSB_j$ for all $j=1, \ldots m$, and $SSW = \sum_{j=1}^m SSW_j$. The within-groups variability for each attribute is defined by
\begin{equation}\label{eqn:c5}
    SSW_j = \displaystyle\sum_{q=1}^k \sum_{i \in Q^q} \left(x_{ij}-\overline{x}_q^j \right)^2\;, \mbox{  for all } j=1, \ldots m
\end{equation}
In this case, the ratio $\frac{SSW}{SST}$ represents the proportion of the total variability still retained in the groups. Besides, the total variance within-groups $SSW$ also characterizes the sum of squares error ($SSE$).

When $k=n$ and $Q^a = \{a\}$ for all $a=1,...,k$, then $R^2 = 1$ because $SSB=SST$, that is, the variance between singletons is the total variance of the data. On the opposite side, if $k=1$ and $Q^1=V$, then $R^2 = 0$ because $SSB=0$, that is, all the variability is in the single component, so there is no between variability among groups. For any other partition $k \in \{2,...,n-1\}$, we have $R^2 \in [0,1]$, because $0 \leq SSB \leq SST$.

Let $\Omega$ represent the entire set os partitions of $V$. For each partition $P \in \Omega$, we denote by $c(P)$ the number of components in $P$ and denote by $R^2(P)$ the $R^2$ ratio of $P$. Considering these notations, the GC problem can be modeled as

\begin{equation}\label{eqn:c6}
    k = \min_{P \in \Omega} \left\{ c(P) : R^2(P) \geq R^2T \right\}
\end{equation}

The $R^2$ ratio provides a global indicator of the total variance captured by the partition, over the entire set of attributes at the same time. It aggregates in a single value the particular variance of each attribute on the same common partition. Yet, if we want to determine each attribute's $R^2$ ratio under the same common partition, then we should use expression (\ref{eqn:c7}).

\begin{equation}\label{eqn:c7}
    \displaystyle R^2_j = \frac{SSB_j}{SST_j} = 1 - \frac{SSW_j}{SST_j}\;, \mbox{  for all } j=1, \ldots m
\end{equation}

We will keep, however, the discussion on the single $R^2$ ratio, as the global condition to be met. Then, we can observe how each attribute $R^2_j$ ratio is represented in the final solution partition.

\begin{figure}
\begin{center}
  \includegraphics[scale=0.35]{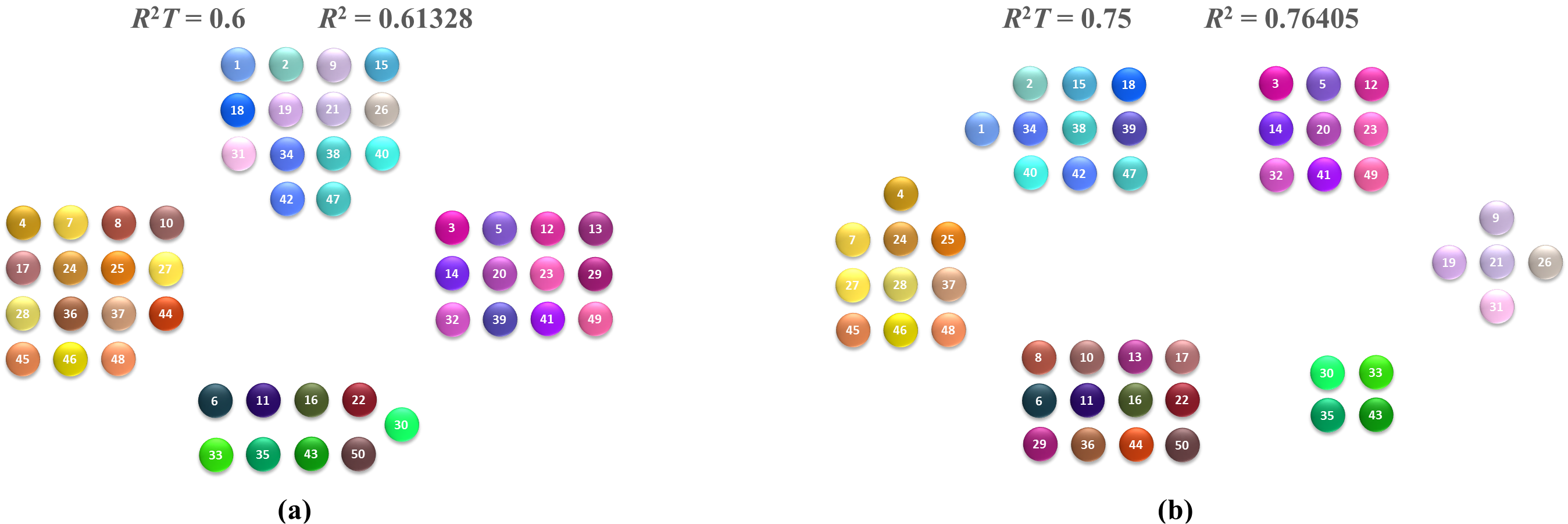}
  \caption{\footnotesize GC optimum solutions with $R^2T=0.6$ (in (a)) and $R^2T=0.75$ (in (b)).}\label{Fig1}
\end{center}
\end{figure}

Figure \ref{Fig1} shows an example involving $n=50$ points on $m=3$ attributes, representing the RGB coordinates of 50 points' colors, randomly generated. The two images characterize two GC optimum solutions, with $R^2T=0.6$ (in (a)) and $R^2T=0.75$ (in (b)).

As expected, the more demanding $R^2T$ threshold provides a more coherent partition (solution in (b)), producing more homogeneous colors in each group. It naturally required more groups to reach the goal. They both used the smallest number of groups reaching the threshold target. The particular $R^2_j$ ratios for each attribute $j=1, \ldots, 3$, that is, for Red ($j=1$), Green ($j=2$) and Blue ($j=3$), are, for the solution in (a): 0.53670, 0.41748 and 0.82217, respectively; and for the solution in (b): 0.75829, 0.68421 and 0.82636, respectively. In this case, attribute 3 (Blue) is, somehow, benefitted, presenting a better $R^2$ ratio. This is not controlled by the model discussed in the present paper. However, it could be an interesting point to explore in a future work, namely to set individual $R^2_j$ threshold ratios.

It is also important to mention that all the discussion in the present paper involves the $R^2$ ratio and not its adjusted version. In effect, this work concentrates on unsupervised non-inferential clustering, which means that we are not analyzing samples but entire data instances, and not expecting additional attributes (variables). Considering the various applications on clustering analysis, we detach, in the aim of the GC problem, those on data reduction, image segmentation, market segmentation, financial markets, gene expression data analysis, among others.

The closest approaches to the GC problem are the Ward's method within hierarchical clustering \cite{War1963, JaiMurFly1999} and the \emph{k}-means algorithm in non-hierarchical clustering \cite{MacQue1967, Jai2010}. These are long known methods and extensively used in practice. They will be described in more detail in Section \ref{sec:Heuristics}. For the moment, however, we simply consider them to compare their answers if used as alternative methods to the GC problem. To this purpose, we resort to the SPSS package, version 24.0.0.

Very briefly, the Ward's method produces an entire sequence of inclusionwised partitions, allowing the user to select a partition by observing the entire hierarchical framework produced. On a different way, the \emph{k}-means finds a partition on a given number of groups ($k$), and then confront the user with the final solution associated $R^2$ ratio (or another indicator). Following an alternative direction, the GC problem sets in advance a $R^2$ ratio threshold and looks for a partition with the minimum number of groups that satisfies the given threshold. To our best knowledge, clustering problems based on Euclidean or squared Euclidean distances have never been posed this way.

Using the example described in Figure \ref{Fig1} and if known in advance that the partitions of interest have a number of groups ($k$) ranging between 3 and 7, we can use the \emph{k}-means (Euclidean version) to find those partitions. The solutions returned have $R^2$ ratios equal to 0.4860, 0.6133, 0.6865, 0.7363 and 0.7977, for $k=3, \ldots, 7$, respectively (using the SPSS package, with the maximum number of iterations set to 999, with the "Iterate and classify" option, with convergence criterion equal to 0, and using running means). Now, to accomplish with the threshold $R^2T=0.6$ the partition with the smallest number of components is the one with $k=4$ and $R^2=0.6133$, agreeing with the solution obtained by the GC problem (Figure \ref{Fig1} (a)). But if the threshold $R^2T=0.75$, then we would need the 7 groups partition, with $R^2=0.7977$, because the one with 6 groups has a $R^2$ ratio smaller than the threshold.

If we use, instead, the Ward's algorithm that also minimizes the sum of squares error (squared Euclidean distances), the partitions with the smaller number of groups and with $R^2$ ratios at least the given thresholds also produce weaker solutions than those obtained by the GC problem. In fact, considering the solutions returned by the Ward's algorithm, for $R^2T=0.6$ the best partition involves 5 groups ($k=5$), with $R^2=0.6695$, because the partition with 4 groups has $R^2=0.5720$; and for $R^2T=0.75$ the best partition has 7 groups ($k=7$), with $R^2=0.7945$, because the 6 groups partition has $R^2=0.7421$.

\begin{figure}
\begin{center}
  \includegraphics[scale=0.35]{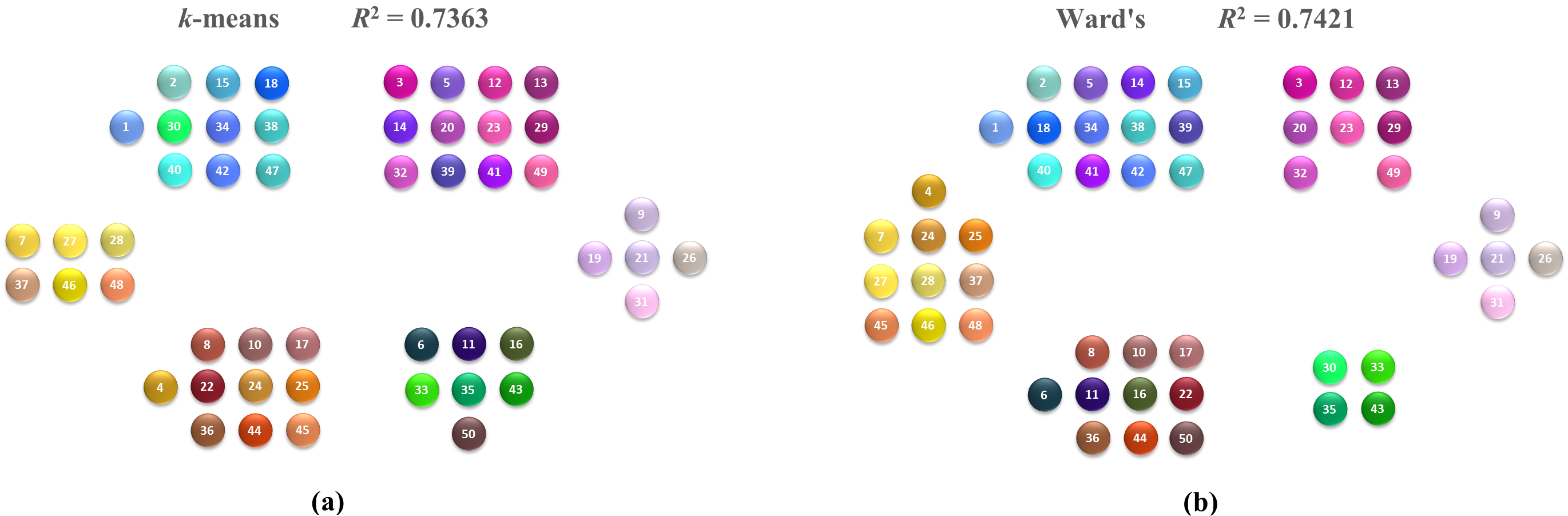}
  \caption{\footnotesize Partitions found by the \emph{k}-means algorithm (in (a)) and by the Ward's method (in (b)) involving 6 groups.}\label{Fig2}
\end{center}
\end{figure}

Figure \ref{Fig2} shows the regretted solutions when the threshold is set to $R^2T=0.75$, produced by the two methods: \emph{k}-means (in (a)) and Ward's (in (b)). These partitions have the same number of components as the solution obtained by the GC problem in Figure \ref{Fig1} (b). According to the $R^2$ ratios involved, the partition produced by the GC should provide more homogeneous groups.

In what follows and in the next section, we propose two VNS based heuristics for the GC problem. Computational tests are conducted in Section \ref{sec:Tests}, considering randomly generated instances and two real-world examples. The first applied example arises from a gene expression microarray dataset and the other one involves a financial markets' datastream. The last section is devoted to conclusions.

\section{Heuristics for the GC problem} \label{sec:Heuristics}

In this section we discuss two VNS based heuristics for the GC problem. The first one uses the Ward's construction to produce the initial approximate solution. The second one performs a similar process but using, instead, a sequence of \emph{k}-means procedures to do the same initial step, that is, to find a starting solution. Before entering the VNS heuristics, we describe the Ward's and the \emph{k}-means based procedures for producing the initial approximate solutions.

\subsection{Ward's algorithm for the GC problem} \label{subsec:W-GC}

Consider a partition $Q^1 \cup Q^2 \cup \ldots  \cup Q^k$ of $V$ into $k$ groups, denoted by $P'$, with $c(P')=k$. We say that a partition $P''$ is neighbor of $P'$ if both partitions have $k-2$ common components and the remaining two are found in $P''$ merged together, that is, $P''$ is obtained from $P'$ by merging two of its components; and thus, $c(P'')=k-1$. The neighborhood of $P'$, denoted by $N(P')$, is the set of all neighbor partitions of $P'$, considering all possible pairs of components' merges. The cardinality of $N(P')$ is $\frac{k \cdot (k-1)}{2}$ and the $R^2$ ratio of any partition in $N(P')$ is not higher than the $R^2$ ratio of $P'$. This property is expressed in Corollary \ref{Cor1}, which is based on the difference between $R^2(P')$ and $R^2(P'')$ for any partition $P'' \in N(P')$, established in Proposition \ref{Prop1}. Its proof is in the Appendix A.

\begin{prop}\label{Prop1}
Given a partition $P'$ of $V$ and its neighborhood $N(P')$. Let $P'' \in N(P')$, with $A$ and $B$ the two merged components from $P'$. Then, $R^2(P') - R^2(P'') = \frac{1}{SST} \cdot \frac{|A|.|B|}{|A|+|B|} \sum_{j=1}^m \left( \overline{x}_A^j  -  \overline{x}_B^j  \right)^2$.
\end{prop}

Considering that the total variance $SST$ is a positive constant for any partition in $\Omega$, then an immediate corollary from Proposition \ref{Prop1} is that $R^2(P') \geq R^2(P'')$, for any $P'' \in N(P')$.

\begin{cor}\label{Cor1}
Given a partition $P'$ of $V$ and its neighborhood $N(P')$. For any $P'' \in N(P')$ we have $R^2(P') \geq R^2(P'')$.
\end{cor}

Yet, it is important to clarify that the result established in Corollary \ref{Cor1} is not extended to partitions that do not verify the hierarchical property. That is, there can exist in $\Omega$ two partitions $P'$ and $P''$ that simply verify the condition $c(P'')=c(P')-1$, and whose $R^2$ ratios do not agree with $R^2(P') \geq R^2(P'')$. This can be observed in some of the results reported in Tables \ref{tab:t2} and \ref{tab:t3} within the computational tests conducted in Subsection \ref{subsec:T-RG}. In effect, the result established in Corollary \ref{Cor1} only holds for $P'' \in N(P')$, that is, if $P''$ is obtained from $P'$ by merging two of its components, or among two partitions $P'$ and $P''$ where $P''$ is obtained from $P'$ by merging some of its components.

However, one way to extend the previous result involves classes of partitions, where in each class we have all partitions with the same number of components. This extension compares the "best" partitions among two classes. To this purpose, let subset $\Omega_i \subset \Omega$ represent the class of all partitions with exactly $i$ components. We denote by $\widehat{P}_i \in \Omega_i$ a partition with the highest $R^2$ ratio among all partitions in $\Omega_i$, that is,

\begin{equation}\label{eqn:c8}
    \widehat{P}_i = \arg \max \left\{ R^2(P) : P \in \Omega_i \right\}
\end{equation}

Considering Corollary \ref{Cor1}, we can state that for any partition $P_i \in \Omega_i$ ($i=2, \ldots, n$), there exists another partition $P_{i-1} \in N(P_i)$ such that $R^2(P_i) \geq R^2(P_{i-1})$, which implies that $R^2(\widehat{P}_i) \geq R^2(\widehat{P}_{i-1})$. This result can be generalized, establishing a hierarchical sequence of maximum $R^2$ ratios' partitions, stated in Corollary \ref{Cor2}.

\begin{cor}\label{Cor2}
Condition $R^2(\widehat{P}_{i}) \geq R^2(\widehat{P}_{i-1})$ holds for any $i \in \{2, \ldots, n\}$.
\end{cor}

The property established in Corollary \ref{Cor2} is only valid among partitions having the highest $R^2$ ratios in their own classes ($\Omega_i$). So, the previous observation comparing pairs of partitions of any kind, ignoring the hierarchical property, still holds. This means that if we take a general pair of partitions from $\Omega$, with different number of components, we cannot guarantee a hierarchical relationship among their $R^2$ ratios.

\bigskip
The agglomerative version of the Ward's algorithm is an iterative process that moves from one partition to another one in its neighborhood, producing a hierarchical sequence of inclusionwise partitions. It starts with $n$ singletons, with $R^2=1$; and ends with all the points in a single component, with $R^2=0$. If $P'$ is the current iteration partition, then the chosen partition $P'' \in N(P')$ is the one with the largest $R^2$ ratio among all partitions in $N(P')$, that is, $P'' = \arg \max \left\{ R^2(P): P \in N(P') \right\}$, producing the smallest $R^2$ deterioration. Let $P^i$ represent the partition determined in iteration $i$, and assume that $P^0=\{1\} \cup \{2\} \cup \ldots \cup \{n\}$. According to Corollary \ref{Cor1}, if we take $P^{i+1} \in N(P^i)$, then $R^2(P^{i}) \geq R^2(P^{i+1})$. Using this fact, the Ward's algorithm can be used to answer the GC problem if its execution is stopped when the current iteration partition has an $R^2$ ratio smaller than the threshold $R^2T$. Thus, if the algorithm stopes at iteration $i$, then $R^2(P^{i+1}) < R^2T$ and $R^2(P^{i}) \geq R^2T$, meaning that $P^{i}$ was the last valid partition in the sequence, having the smallest number of groups among all valid partitions found by the Ward's algorithm. The number of groups in $P^{i}$ is $c(P^{i})=n-i$, providing an approximate answer to the GC problem. Figure \ref{Fig3} describes the algorithm, with parameter $R^2T$ as the single input.

\begin{figure}[h]
\begin{center}
\begin{tabular}{|ll|}
\hline
  & \texttt{\textbf{procedure Ward's-GC}} \\
1 & Let $P^0=\{1\} \cup \{2\} \cup \ldots \cup \{n\}$ and set $i \leftarrow 0$ \\
2 & \textbf{while} $R^2(P^i) \geq R^2T$ \textbf{do} \\
3 & \hspace*{0.5cm} Determine $P^{i+1} = \arg \max \left\{ R^2(P): P \in N(P^i) \right\}$ \\
4 & \hspace*{0.5cm} Make $i \leftarrow i+1$ \\
5 & \textbf{end-while} \\
6 & Return $P^{i-1}$ and $c(P^{i-1})$ \\
\hline
\end{tabular}
\caption{\label{Fig3} Ward's algorithm for the GC problem.}
\end{center}
\end{figure}

Considering the 50 points (colors) example used in Section \ref{sec:Intro}, the solution produced by the Ward's-GC algorithm for a threshold $R^2T=0.75$ is the partition shown in Figure \ref{Fig2} (b).

Note that maximizing the $R^2(P)$ ratio (used in the expression in line 3 of the algorithm) is the same as minimizing the within-groups variability ($SSW$), according to expression (\ref{eqn:c4}); and the same as maximizing the between-groups variability ($SSB$), according to expression (\ref{eqn:c1}); because the total variability ($SST$) is a constant for any partition of $V$.

A known drawback of the Ward's procedure is that the partition found in each iteration $i$ may not represent the partition with highest $R^2$ ratio among all partitions on $n-i$ groups. This is due to the inclusionwise property of the algorithm that restricts the selection performed in iteration $i$ to a very tiny region of the space of all partitions on $n-i$ groups. In fact, it only looks to the neighborhood $N(P^i)$.

In view of the forthcoming discussion involving the VNS based algorithms, we can describe a different version of the Ward's procedure by substituting the starting solution $P^0$ by an advanced partition $P'$, allowing the procedure to start at a later stage. This version is denoted by \textbf{Ward's-GC($P'$)}, and it can be obtained from the former by substituting line 1 by "Set $i \leftarrow c(P')$ and $P^i \leftarrow P'$". The new starting partition should satisfy the condition $R^2(P') \geq R^2T$.

\subsection{\emph{k}-means based process for the GC problem} \label{subsec:kmeans-GC}

An alternative methodology for solving (heuristically) the GC problem resorts to the \emph{k}-means algorithm. The \emph{k}-means is a non-hierarchical clustering method that finds a partition with a fixed number of components ($k$). It starts with an initial partition with $k$ components and computes the associated centroids. Then, all points are reassigned to their closest centroid, producing a new partition; and the centroids are readjusted once again. The algorithm iterates until no point is replaced. Closeness is calculated using the Euclidean (or squared Euclidean) distance between each point and its own component centroid. The version here discussed uses the Euclidean distance form. Figure \ref{Fig4} describes the \emph{k}-means algorithm for a given number of components $k$.

\begin{figure}[h]
\begin{center}
\begin{tabular}{|ll|}
\hline
  & \texttt{\textbf{procedure \emph{k}-means($k$)}} \\
1 & Let $P^0$ be an initial partition on $k$ components and set $i \leftarrow 0$ \\
2 & Determine each component centroid \\
4 & \textbf{repeat} \\
5 & \hspace*{0.5cm} Make $i \leftarrow i + 1$ \\
6 & \hspace*{0.5cm} Reassign each point to its closest centroid, leading to $P^i$ \\
7 & \hspace*{0.5cm} Readjust the centroids in the modified components \\
9 & \textbf{until} there are no reassignments \\
10 & Return $P^{i-1}$ \\
\hline
\end{tabular}
\caption{\label{Fig4} The \emph{k}-means algorithm.}
\end{center}
\end{figure}

The \emph{k}-means is an NP-hard problem, even for $k=2$ \cite{Drietal1999}. The algorithm described in Figure \ref{Fig4} is a heuristic process that finds an approximate solution to the partitioning problem into $k$ components, minimizing the total distances from each point to its own component centroid. So, it converges to a local minimum. In addition, the \emph{k}-means is strongly sensitive to the initial partition or the initial seeds, as observed by many authors (see, e.g., \cite{SelIsm1984}, \cite{BabMur1993}, \cite{JaiMurFly1999}, \cite{RubSouUgo2006} and \cite{Festa2013}). There are various suggestions for handling this starting step, namely running that algorithm several times starting from different partitions. In our case, however, we used the \emph{k}-medoids algorithm for furnishing the initial partition. The \emph{k}-medoids is very similar to the \emph{k}-means algorithm, producing closely related optimum solutions. The major difference is that the seeds in the \emph{k}-medoids are taken from the set of points ($V$), while in the \emph{k}-means the seeds are any \emph{m}-dimensional vector in the $R^m$ space. The \emph{k}-medoids is a long known problem within the Combinatorial Optimization community, designated by \emph{p}-median problem (where $p=k$). It also belongs to the NP-hard class \cite{KarHak1979} and its applicability has been concentrating the attention of many researchers from various scientific domains.

For the purpose of generating an initial partition, we use a composite heuristic \cite{Mlaetal2007} for the \emph{p}-median problem, combining the greedy procedure proposed in \cite{KueHam1963} with a simple local search process. The greedy procedure builds one component at a time, starting with the 1-median solution, and opening a new component in each iteration, until obtaining the \emph{p} components. Each new component is centralized in a new element, taken among non-central candidates, and such that the total cost is minimized. The final solution of the greedy algorithm is used to start a local search phase. This phase uses the elements that were not selected to become centers during the greedy execution, but that could further reduce the total cost if the greedy procedure has not been stopped. Those non central elements are introduced in a list. This list of candidate elements is used for local permutations, substituting a selected central element in the current solution by an element in the list; and reassigning the elements according to the new pool of central elements. The new solution becomes the incumbent if its cost is lower than the former. This interchange process is performed until stabilizing the current incumbent. The entire process (greedy procedure plus local search) is very fast, constituting an efficient procedure to generate the required initial solution for the \emph{k}-means.

Considering the \emph{k}-means algorithm described in Figure \ref{Fig4}, with the mentioned \emph{p}-median composite heuristic for setting the initial partition, we propose using a bisection methodology for building a new algorithm for the GC problem. It starts with the set $\{a, \ldots, b\}$, with $a=1$ and $b=n$, representing all possible number of components in a partition. Let $P^a$ and $P^b$ denote the partitions returned by the \emph{k}-means algorithm, for $k=a$ and $k=b$, respectively, with $R^2(P^a) < R^2T$ and $R^2(P^b) > R^2T$, assuming that $R^2T \neq 0$ and $R^2T \neq 1$. Calculate the mid integer in the range $\{a, \ldots, b\}$, determined by $c=\lfloor\frac{a+b}{2}\rfloor$, and call procedure \emph{k}-means(\emph{c}), generating a new partition $P^c$. If $R^2(P^c)=R^2T$, then the algorithm stops, returning $c$ as the number of components in the best partition found, and whose $R^2$ ratio respects the threshold. Otherwise, if $R^2(P^c)<R^2T$, then the lowest number of components in a partition that satisfies the threshold is in the set $\{c, \ldots, b\}$, and we make $a \leftarrow c$; or, if $R^2(P^c)>R^2T$, the the lowest number of components in a partition that respects the threshold is in the set $\{a, \ldots, c\}$, and we make $b \leftarrow c$. The new range $\{a, \ldots, b\}$ still verifies the conditions $R^2(P^a) < R^2T$ and $R^2(P^b) > R^2T$. It is once again divided in two subsets, from which we take the half that contains the partition with the lowest number of components, whose $R^2$ ratio satisfies the threshold. The algorithm iterates this way, halving the current set $\{a, \ldots, b\}$ and taking the half that contains the appropriate number of components. It stops when $(b-a) < 2$, taking $b$ as the best solution found, representing an approximate solution to problem GC. This algorithm is described in Figure \ref{Fig5}, also with parameter $R^2T$ as the single input.

\begin{figure}[h]
\begin{center}
\begin{tabular}{|ll|}
\hline
  & \texttt{\textbf{procedure \emph{k}means-GC}} \\
1 & Make $a=1$ and $b=n$ \\
2 & \textbf{while} $(b-a) \geq 2$ \textbf{do}\\
3 & \hspace*{0.5cm} Determine $c=\lfloor\frac{a+b}{2}\rfloor$ \\
4 & \hspace*{0.5cm} Call \emph{k}-means(\emph{c}), returning partition $P^c$\\
5 & \hspace*{0.5cm} Calculate $R^2(P^c)$ \\
6 & \hspace*{0.5cm} \textbf{if} $R^2(P^c) = R^2T$ \textbf{then}\\
7 & \hspace*{1.0cm} Make $a \leftarrow c$ and $b \leftarrow c$ \\
8 & \hspace*{0.5cm} \textbf{else} \\
9 & \hspace*{1.0cm} \textbf{if} $R^2(P^c) < R^2T$ \textbf{then} \\
10 & \hspace*{1.4cm} Make $a \leftarrow c$ \\
11 & \hspace*{1.0cm} \textbf{else} \\
12 & \hspace*{1.4cm} Make $b \leftarrow c$ \\
13 & \hspace*{1.0cm} \textbf{end-if} \\
14 & \hspace*{0.5cm} \textbf{end-if} \\
15 & \textbf{end-while} \\
16 & Return $P^b$ and $c(P^b)$ \\
\hline
\end{tabular}
\caption{\label{Fig5} \emph{k}-means based algorithm for the GC problem.}
\end{center}
\end{figure}


\subsection{VNS based algorithms} \label{subsec:VNS}

The VNS metaheuristic has long been used to address many combinatorial optimization problems with large success, including clustering related problems (see, e.g., \cite{CafHanMla2014}, \cite{Caretal2013}, \cite{ConSti2015}, \cite{ZhiRus2015}). Two survey works on the VNS can be found in \cite{HanMla2003} and \cite{HanMlaPer2010}. VNS basic schemes and variants are discussed in \cite{Hanetal2016}, including implementation aspects and hybrid versions with other metaheuristics.

The VNS is based on a systematic change of neighborhoods in searching the solutions space of a combinatorial optimization problem. In this paper we follow the basic version of the VNS, being used for improving the solutions returned by each of the GC algorithms described in Subsections \ref{subsec:W-GC} and \ref{subsec:kmeans-GC}.

To this purpose, we introduce a new neighborhood concept. Given a partition $P \in \Omega$, the $r^{\footnotesize \mbox{th}}$ neighborhood of \emph{P}, denoted by $V_r(P)$, is the set of all partitions $P'$ obtained from \emph{P} after removing \emph{r} elements from the current components and putting them in isolated components, and such that none of the current components becomes empty. This means that the structural difference between $P$ and any $P' \in V_r(P)$ are $c(P')-c(P)=r$ components. These new components are singletons, containing the elements that were displaced from the original components of $P$.

This algorithm is described in Figure \ref{Fig6}. It starts calling procedure \textbf{H} for determining an initial solution for the GC problem. This procedure can be any of the former algorithms described in Subsections \ref{subsec:W-GC} or \ref{subsec:kmeans-GC}. If H is procedure Ward's-GC, then the associated VNS is denoted by \textbf{VNS/Ward's-GC}; while if H is \emph{k}means-GC, then the VNS is denoted by \textbf{VNS/\emph{k}means-GC}.
The inputs of the VNS/H are the threshold $R^2T$ ratio, and the limit ($r_{\footnotesize\mbox{max}}$) that establishes the maximum number of elements to remove from the current components in \emph{P}, performed in the shaking phase. This limit should respect the range $1 \leq r_{\footnotesize\mbox{max}} < n - c(P^*)$. In effect, parameter $r_{\footnotesize\mbox{max}}$ is chosen such that $r_{\footnotesize\mbox{max}} \ll n - c(P^*)$.

\begin{figure}[h]
\begin{center}
\begin{tabular}{|ll|}
\hline
  & \texttt{\textbf{procedure VNS/H}} \\
1 & Call procedure \textbf{H}, returning an initial partition $P$ and set $P^* \leftarrow P$ \\
2 & Set $r \leftarrow 1$ and define $r_{\footnotesize\mbox{max}}$ such that $1 \leq r_{\footnotesize\mbox{max}} < n - c(P^*)$ \\
3 & \textbf{while} $r \leq r_{\footnotesize\mbox{max}}$ \textbf{do}\\
4 & \hspace*{0.5cm} \underline{Shaking}: Obtain $P' \in V_r(P^*)$ \\
5 & \hspace*{0.5cm} \underline{Improvement}: Call Ward's-GC($P'$), returning partition $P''$ \\
6 & \hspace*{0.5cm} \textbf{if} $\left( c(P'') < c(P^*) \right)$ or $\left( c(P'') = c(P^*) \mbox{ and } R^2(P'') > R^2(P^*) \right)$ \textbf{then}\\
7 & \hspace*{1.0cm} Make $P^* \leftarrow P''$ and $r \leftarrow 1$ \\
8 & \hspace*{0.5cm} \textbf{else} \\
9 & \hspace*{1.0cm} Make $r \leftarrow r + 1$ \\
10 & \hspace*{0.5cm} \textbf{end-if} \\
11 & \textbf{end-while} \\
12 & Return $P^*$ and $c(P^*)$ \\
\hline
\end{tabular}
\caption{\label{Fig6} VNS based algorithm for the GC problem.}
\end{center}
\end{figure}

In each iteration, a new partition $P'$ is taken from the neighborhood $V_r(P^*)$. This is done by removing $r$ elements from the components characterizing $P^*$, originating $r$ singletons, and such that none of the original components becomes empty, which guarantees that $c(P') = c(P) + r$. The removed elements are chosen in the following way:

\begin{enumerate}
  \item Determine the "removing effect" for all candidate elements
  \item Select the $r$ elements with the highest "removing effect"
\end{enumerate}

Concerning point 1: Given a partition $P$. Take one of its components (excluding singletons), say component A, and remove from A one of its elements, say $b \in A$, originating the singleton composed by $b$. Let $P'$ be the newly formed partition. Now, recovering Proposition \ref{Prop1}, we have that $R^2(P') - R^2(P) = \frac{1}{SST} \cdot \frac{|A|-1}{|A|} \sum_{j=1}^m \left( \overline{x}_{A\setminus\{b\}}^j - x_{bj}  \right)^2$, that is, $R^2(P') - R^2(P) = \frac{1}{SST} \cdot \frac{|A|}{|A|-1} \sum_{j=1}^m \left( \overline{x}_A^j - x_{bj}  \right)^2$, representing the "removing effect" considered in point 1.

Concerning point 2: The selection process performed in this point is accompanied by a random strategy. So, consider the sorted list of the candidate elements, organized according to the "removing effect", from highest to lowest. Following this list, the $i^{\footnotesize\mbox{th}}$ candidate is selected (removed from the current component) only if $\rho > \frac{i}{\min \left\{ n, 2 r \right\}}$, for $\rho$ a randomly generated value following the uniform distribution $U(0,1)$. This random effect interferes less on the first elements in the list, allowing a stronger chance for them to be selected; while the last elements have a lower chance. This strategy promotes the elements with higher "removing effect", concentrating the diversification mechanisms in the elements with lower "removing effect". In order to guarantee the effective selection of all the $r$ elements, we set parameter $r_{\footnotesize\mbox{max}} \ll n - c(P^*)$.

Then, the newly generated partition $P' \in V_r(P^*)$ is used to start a new constructing phase performed by the Ward's-GC($P'$), returning partition $P''$. If the newly obtained partition $P''$ is better than the current best solution ($P^*$), then $P^*$ is updated and $r$ is resumed, contracting the neighborhood radius; otherwise, parameter $r$ is increased and the neighborhood radius grows. This process stops when the neighborhood radius ($r$) reaches $r_{\footnotesize\mbox{max}}$. In addition, we say that partition $P''$ is better than $P^*$ if $P''$ has less components than $P^*$, that is, $c(P'') < c(P^*)$; or, if $P''$ has a better $R^2$ ratio than $P^*$ in the case where the two partitions have the same number of components, that is, when $c(P'') = c(P^*) \mbox{ and } R^2(P'') > R^2(P^*)$.

\section{Computational tests} \label{sec:Tests}

In order to test the computational performance of the GC algorithms proposed in Section \ref{sec:Heuristics}, we provide some computational experiments using real-world and randomly generated instances.
All tests were conducted on an Intel Core i7-2600 with 3.40 GHz and 8 GB RAM. The experiments were performed under Microsoft Windows 10 operating system. The algorithms described in Figures \ref{Fig3} to \ref{Fig6} were coded in Fortran, and compiled on gfortran.

An important aspect to detach in the implementation of the two procedures that generate initial solutions for the VNS algorithms is the memory size that they require. In these implementations, the Ward's-GC requires $O(nm)$ memory space, while the \emph{k}means-GC requires $O(n^2)$, if $n \geq m$ and $O(nm)$ if $n<m$. This aspect is key to handle very large sized instances as showed in the computational tests here discussed, besides the execution complexity of the two algorithms.
The reason for the $O(n^2)$ memory requested by the \emph{k}means-GC is related with the \emph{p}-median implementation, used for generating the initial partition for the \emph{k}-means procedure. We could substitute the \emph{p}-median algorithm by an alternative less demanding procedure; or redesign the \emph{p}-median in order to produce an implementation requiring just $O(mn)$ memory space. However, those alternative versions could possibly decrease the quality of the \emph{k}-means solutions, or impose a higher execution time, considering the relevancy of this initial step in the \emph{k}-means performance, detached by many authors in the literature.

The maximum number of iterations of the VNS was set to 50, that is, we set $k_{\mbox{max}}=50$ in all the tests conducted in the present paper.

We consider two types of instances. The first one involves randomly generated instances, while the second one uses real-world datasets. All datasets have been standardized, which means that for each attribute, the entries' mean is 0 and the standard-deviation is 1.

\subsection{Tests on the randomly generated instances} \label{subsec:T-RG}

The randomly generated instances involve two classes. In the first one, the entries follow the standardized Normal distribution, meaning that $v(i,j) \sim N(0,1)$, for all $i=1, \ldots, n$ and $j=1, \dots, m$; while in the second one, the entries follow an Uniform distribution, taking values in the interval $[-1,1]$, that is, $v(i,j) \sim U(-1,1)$, for all $i=1, \ldots, n$ and $j=1, \dots, m$. Various instances, with different sizes, have been generated for both classes. The number of elements in these instances take the values $n=100, 500 \mbox{ and } 1000$, with $m=3, 0.01n, 0.05n, 0.1n$, for setting the number of attributes. Table \ref{tab:t1} shows the various combinations of the two parameters, $n$ and $m$, that characterize each instance. Each entry in the table represents an instance, and the value indicates the associated number of attributes. A blank cell indicates that the associated instance was not generated.

\smallskip
{\small
\begin{table}[h]
  \centering
 \begin{tabular}{l|cccc}
     & $m=3$ & $m=0.01n$ & $m=0.05n$ & $m=0.1n$ \\
  \hline
  $n=100$ & 3 & & 5 & 10 \\
  $n=500$ & 3 & 5 & 25 & 50 \\
  $n=1000$ & 3 & 10 & 50 & 100 \\
\end{tabular}
  \caption{\small Number of elements ($n$) and number of attributes ($m$) in the first set of randomly generated instances.}\label{tab:t1}
\end{table}}

The instances generated using the standard Normal distribution $N(0,1)$ are denoted by N-\emph{m}-\emph{n}, and those generated using the Uniform distribution $U(-1,1)$ are denoted by U-\emph{m}-\emph{n}.

Table \ref{tab:t2} presents the results of the GC algorithms proposed in Section \ref{sec:Heuristics} considering the instances generated using the Normal distribution, while Table \ref{tab:t3} presents the results returned by the same algorithms considering the instances following the Uniform distribution. Each instance was tested for three values of the threshold parameter, considering $R^2T = 0.6, 0.7 \mbox{ and } 0.8$. For each algorithm we report the number of components in the best partition ($c(P)$), the associated $R^2$ ratio ($R^2(P)$), and the execution time (in seconds). The execution time of the algorithms VNS/Ward's-GC and VNS/\emph{k}means-GC include the time to determine the first partition to start the VNS, performed by the algorithms Ward's-GC and \emph{k}means-GC, respectively. The values in \textbf{bold} indicate the best solutions found for the associated instance; and we use the notation '$< 1$' whenever the execution time is less than 1 second. The maximum execution time was set to 21600 seconds (6 hours).

\smallskip {\small
\begin{table}
  \centering
 {\footnotesize
 \begin{tabular}{lc|ccc|ccc|ccc|ccc|}
  & & \multicolumn{3}{c}{Ward's-GC} & \multicolumn{3}{c|}{VNS/Ward's-GC} & \multicolumn{3}{|c}{\emph{k}means-GC} & \multicolumn{3}{c|}{VNS/\emph{k}means-GC}
    \\ 
  Instances & $R^2T$ & $c(P)$ & $R^2(P)$ & time & $c(P)$ & $R^2(P)$ & time & $c(P)$ & $R^2(P)$ & time & $c(P)$ & $R^2(P)$ & time \\
  \hline
  N-100-3 & 0.6 & 6 & 0.623 & $< 1$ & 6 & 0.635 & $< 1$ & 6 & 0.634 & $< 1$ & \textbf{6} & \textbf{0.639} & $< 1$ \\
  N-100-3 & 0.7 & 9 & 0.726 & $< 1$ & 8 & 0.713 & $< 1$ & \textbf{8} & \textbf{0.717} & $< 1$ & \textbf{8} & \textbf{0.717} & $< 1$ \\
  N-100-3 & 0.8 & 13 & 0.802 & $< 1$ & \textbf{12} & \textbf{0.807}  & $< 1$ & 13 & 0.809 & $< 1$ & 12 & 0.806 & $< 1$ \\
  \hline
  N-100-5 & 0.6 & 9 & 0.614 & $< 1$ & \textbf{9} & \textbf{0.625} & $< 1$ & 9 & 0.617 & $< 1$ & \textbf{9} & \textbf{0.625} & $< 1$ \\
  N-100-5 & 0.7 & 13 & 0.707 & $< 1$ & \textbf{13} & \textbf{0.709} & $< 1$ & 14 & 0.711 & $< 1$ & \textbf{13} & \textbf{0.709} & $< 1$ \\
  N-100-5 & 0.8 & 21 & 0.804 & $< 1$ & 21 & 0.806 & $< 1$ & 24 & 0.807 & $< 1$ & \textbf{21} & \textbf{0.807} & $< 1$ \\
  \hline
  N-100-10 & 0.6 & 21 & 0.612 & $< 1$ & \textbf{20} & \textbf{0.608} & 3 & 24 & 0.601 & $< 1$ & 20 & 0.602 & 3 \\
  N-100-10 & 0.7 & 29 & 0.700 & $< 1$ & 29 & 0.702 & 3 & 35 & 0.703 & $< 1$ & \textbf{29} & \textbf{0.705} & 10 \\
  N-100-10 & 0.8 & 43 & 0.803 & $< 1$ & 43 & 0.804 & 3 & 48 & 0.802 & $< 1$ & \textbf{43} & \textbf{0.806} & 5 \\
  \hline
  N-500-3 & 0.6 & 8 & 0.614 & 1 & 7 & 0.604 & 4 & \textbf{6} & \textbf{0.606} & 9 & \textbf{6} & \textbf{0.606} & 10 \\
  N-500-3 & 0.7 & 12 & 0.706 & 1 & 11 & 0.701 & 3 & \textbf{11} & \textbf{0.713} & 9 & \textbf{11} & \textbf{0.713} & 11 \\
  N-500-3 & 0.8 & 21 & 0.804 & 1 & 19 & 0.800 & 6 & 19 & 0.802 & 10 & \textbf{19} & \textbf{0.803} & 10 \\
  \hline
  N-500-5 & 0.6 & 18 & 0.609 & 2 & 17 & 0.610 & 8 & \textbf{15} & \textbf{0.606} & 13 & \textbf{15} & \textbf{0.606} & 15 \\
  N-500-5 & 0.7 & 30 & 0.703 & 2 & 29 & 0.703 & 7 & \textbf{26} & \textbf{0.703} & 13 & \textbf{26} & \textbf{0.703} & 16 \\
  N-500-5 & 0.8 & 54 & 0.801 & 2 & 52 & 0.800 & 13 & 54 & 0.802 & 15 & \textbf{52} & \textbf{0.803} & 29 \\
  \hline
  N-500-25 & 0.6 & 136 & 0.601 & 18 & \textbf{134} & \textbf{0.601} & 859 & 165 & 0.601 & 30 & 138 & 0.601 & 1630 \\
  N-500-25 & 0.7 & \textbf{194} & \textbf{0.701} & 17 & \textbf{194} & \textbf{0.701} & 409 & 228 & 0.701 & 45 & \textbf{194} & \textbf{0.701} & 1412 \\
  N-500-25 & 0.8 & 269 & 0.801 & 15 & 269 & 0.801 & 197 & 299 & 0.800 & 58 & \textbf{268} & \textbf{0.800} & 2441 \\
  \hline
  N-500-50 & 0.6 & 188 & 0.602 & 41 & \textbf{187} & \textbf{0.601} & 907 & 225 & 0.601 & 38 & 189 & 0.600 & 5813 \\
  N-500-50 & 0.7 & \textbf{247} & \textbf{0.700} & 36 & \textbf{247} & \textbf{0.700} & 526 & 284 & 0.701 & 57 & 249 & 0.701 & 4052 \\
  N-500-50 & 0.8 & \textbf{318} & \textbf{0.801} & 29 & \textbf{318} & \textbf{0.801} & 502 & 348 & 0.800 & 45 & 318 & 0.800 & 2457 \\
  \hline
  N-1000-3 & 0.6 & 8 & 0.621 & 9 & 8 & 0.628 & 12 & \textbf{6} & \textbf{0.602} & 258 & \textbf{6} & \textbf{0.602} & 260 \\
  N-1000-3 & 0.7 & 13 & 0.712 & 9 & 12 & 0.710 & 14 & \textbf{10} & \textbf{0.704} & 261 & \textbf{10} & \textbf{0.704} & 263 \\
  N-1000-3 & 0.8 & 23 & 0.803 & 9 & 22 & 0.805 & 17 & \textbf{19} & \textbf{0.806} & 262 & \textbf{19} & \textbf{0.806} & 265 \\
  \hline
  N-1000-10 & 0.6 & 86 & 0.601 & 46 & 85 & 0.602 & 162 & 79 & 0.600 & 201 & \textbf{79} & \textbf{0.601} & 256 \\
  N-1000-10 & 0.7 & 151 & 0.701 & 46 & 148 & 0.700 & 398 & 156 & 0.700 & 234 & \textbf{148} & \textbf{0.701} & 740 \\
  N-1000-10 & 0.8 & 262 & 0.800 & 43 & \textbf{260} & \textbf{0.800} & 1165 & 289 & 0.800 & 449 & 265 & 0.800 & 3429 \\
  \hline
  N-1000-50 & 0.6 & \textbf{358} & \textbf{0.601} & 331 & \textbf{358} & \textbf{0.601} & 4597 & 435 & 0.601 & 543 & 362 & 0.600 & 21600 \\
  N-1000-50 & 0.7 & \textbf{479} & \textbf{0.700} & 290 & \textbf{479} & \textbf{0.700} & 3850 & 553 & 0.700 & 787 & 483 & 0.700 & 21600 \\
  N-1000-50 & 0.8 & \textbf{622} & \textbf{0.800} & 248 & \textbf{622} & \textbf{0.800} & 3249 & 683 & 0.800 & 791 & 626 & 0.800 & 21600 \\
  \hline
  N-1000-100 & 0.6 & 435 & 0.600 & 673 & \textbf{435} & \textbf{0.601} & 10171 & 504 & 0.600 & 896 & 441 & 0.600 & 21600 \\
  N-1000-100 & 0.7 & \textbf{554} & \textbf{0.700} & 586 & \textbf{554} & \textbf{0.700} & 9250 & 614 & 0.701 & 790 & 558 & 0.701 & 21600 \\
  N-1000-100 & 0.8 & \textbf{686} & \textbf{0.800} & 470 & \textbf{686} & \textbf{0.800} & 6965 & 733 & 0.801 & 702 & 689 & 0.800 & 21600 \\
  \hline
\end{tabular}}
  \caption{{\small Results from the proposed GC heuristics for the Normal distributed instances with $n \leq 1000$.}} \label{tab:t2}
\end{table}}

Considering the results reported in Table \ref{tab:t2}, and looking to the Ward's based algorithms, the VNS has been able to decrease the number of components in nearly half of the instances, removing up to 3 components; while on the \emph{k}-means based algorithms, the VNS reduction is much more pronounced, basically because the starting solution produced by the \emph{k}means-GC is very weak, especially for the larger sized instances (large number of elements (\emph{n}) and large number of attributes (\emph{m})). Now, comparing the results among the two VNS approaches, we can observe that the VNS/\emph{k}means-GC can produce solutions with a smaller number of components on the instances with fewer attributes; while the VNS/Ward's-GC performs better on the instances with a larger number of attributes. However, in most cases, the execution time of the VNS/\emph{k}means-GC is much more demanding, especially when the number of elements (\emph{n}) increase. This observation can justify the low performance of the VNS/\emph{k}means-GC on the instances N-1000-50 and N-1000-100, which were interrupted prematurely due to the time limit imposed.

The results also confirm an observation stressed in Subsection \ref{subsec:W-GC}, right after Corollary \ref{Cor1}, stating that for any pair of partitions $P'$ and $P''$ from $\Omega$, with $c(P')>c(P'')$, it may happen that $R^2(P') < R^2(P'')$, namely when $P'' \notin N(P')$. This can be observed, for example in Table \ref{tab:t2}, in instance N-100-3 with $R^2T=0.8$, among the results of the two algorithms, Ward's-GC and VNS/Ward's-GC, with solutions $P^a$ and $P^b$, respectively, where $c(P^a)=13>12=c(P^b)$ and $R^2(P^a) = 0.802 < 0.807 = R^2(P^b)$.

\smallskip {\small
\begin{table}
  \centering
 {\footnotesize
 \begin{tabular}{lc|ccc|ccc|ccc|ccc|}
  & & \multicolumn{3}{c}{Ward's-GC} & \multicolumn{3}{c|}{VNS/Ward's-GC} & \multicolumn{3}{|c}{\emph{k}means-GC} & \multicolumn{3}{c|}{VNS/\emph{k}means-GC}
    \\ 
  Instances & $R^2T$ & $c(P)$ & $R^2(P)$ & time & $c(P)$ & $R^2(P)$ & time & $c(P)$ & $R^2(P)$ & time & $c(P)$ & $R^2(P)$ & time \\
  \hline
  U-100-3 & 0.6 & 5 & 0.643 & $< 1$ & 5 & 0.656 & $< 1$ & 5 & 0.650 & $< 1$ & \textbf{5} & \textbf{0.658} & $< 1$ \\
  U-100-3 & 0.7 & 6 & 0.706 & $< 1$ & \textbf{6} & \textbf{0.722} & $< 1$ & 6 & 0.720 & $< 1$ & 6 & 0.720 & $< 1$ \\
  U-100-3 & 0.8 & 10 & 0.819 & $< 1$ & 9 & 0.806 & $< 1$ & 10 & 0.823 & $< 1$ & \textbf{9} & \textbf{0.808} & $< 1$ \\
  \hline
  U-100-5 & 0.6 & 9 & 0.610 & $< 1$ & 9 & 0.622 & $< 1$ & 9 & 0.625 & $< 1$ & \textbf{8} & \textbf{0.603} & $< 1$ \\
  U-100-5 & 0.7 & 13 & 0.709 & $< 1$ & \textbf{12} & \textbf{0.705} & $< 1$ & 13 & 0.717 & $< 1$ & 12 & 0.704 & $< 1$ \\
  U-100-5 & 0.8 & 20 & 0.808 & $< 1$ & \textbf{19} & \textbf{0.809} & $< 1$ & 20 & 0.805 & $< 1$ & 19 & 0.807 & $< 1$ \\
  \hline
  U-100-10 & 0.6 & 20 & 0.608 & $< 1$ & \textbf{19} & \textbf{0.608} & 2 & 21 & 0.606 & $< 1$ & 19 & 0.607 & 2 \\
  U-100-10 & 0.7 & 29 & 0.706 & $< 1$ & \textbf{28} & \textbf{0.702} & 2 & 32 & 0.703 & $< 1$ & \textbf{28} & \textbf{0.702} & 3 \\
  U-100-10 & 0.8 & 42 & 0.804 & $< 1$ & 42 & 0.805 & 2 & 45 & 0.807 & $< 1$ & \textbf{41} & \textbf{0.801} & 3 \\
  \hline
  U-500-3 & 0.6 & 6 & 0.642 & 1 & 6 & 0.676 & 4 & \textbf{5} & \textbf{0.627} & 10 & \textbf{5} & \textbf{0.627} & 10 \\
  U-500-3 & 0.7 & 8 & 0.736 & 1 & 7 & 0.714 & 3 & \textbf{7} & \textbf{0.729} & 10 & \textbf{7} & \textbf{0.729} & 10 \\
  U-500-3 & 0.8 & 12 & 0.804 & 1 & 12 & 0.811 & 3 & \textbf{11} & \textbf{0.807} & 10 & \textbf{11} & \textbf{0.807} & 11 \\
  \hline
  U-500-5 & 0.6 & 14 & 0.614 & 2 & 12 & 0.607 & 7 & \textbf{11} & \textbf{0.607} & 12 & \textbf{11} & \textbf{0.607} & 14 \\
  U-500-5 & 0.7 & 21 & 0.701 & 2 & 19 & 0.707 & 9 & \textbf{17} & \textbf{0.702} & 12 & \textbf{17} & \textbf{0.702} & 15 \\
  U-500-5 & 0.8 & 40 & 0.802 & 2 & \textbf{36} & \textbf{0.801} & 15 & 37 & 0.803 & 13 & \textbf{36} & \textbf{0.801} & 17 \\
  \hline
  U-500-25 & 0.6 & 135 & 0.601 & 16 & \textbf{133} & \textbf{0.601} & 264 & 159 & 0.601 & 25 & 134 & 0.601 & 881 \\
  U-500-25 & 0.7 & 192 & 0.700 & 15 & 192 & 0.701 & 746 & 222 & 0.700 & 37 & \textbf{191} & \textbf{0.700} & 1947 \\
  U-500-25 & 0.8 & \textbf{268} & \textbf{0.801} & 12 & \textbf{268} & \textbf{0.801} & 188 & 293 & 0.801 & 44 & 269 & 0.801 & 1025 \\
  \hline
  U-500-50 & 0.6 & 188 & 0.601 & 33 & \textbf{187} & \textbf{0.601} & 1307 & 220 & 0.600 & 34 & 189 & 0.602 & 3479 \\
  U-500-50 & 0.7 & \textbf{249} & \textbf{0.701} & 29 & \textbf{249} & \textbf{0.701} & 555 & 283 & 0.700 & 46 & 251 & 0.701 & 3319 \\
  U-500-50 & 0.8 & \textbf{321} & \textbf{0.801} & 22 & \textbf{321} & \textbf{0.801} & 495 & 345 & 0.800 & 42 & 321 & 0.800 & 2651 \\
  \hline
  U-1000-3 & 0.6 & 6 & 0.608 & 9 & 6 & 0.617 & 15 & \textbf{5} & \textbf{0.610} & 204 & \textbf{5} & \textbf{0.610} & 205 \\
  U-1000-3 & 0.7 & 9 & 0.724 & 9 & 8 & 0.739 & 22 & \textbf{7} & \textbf{0.715} & 209 & \textbf{7} & \textbf{0.715} & 211 \\
  U-1000-3 & 0.8 & 15 & 0.810 & 9 & 13 & 0.806 & 17 & \textbf{12} & \textbf{0.809} & 205 & \textbf{12} & \textbf{0.809} & 207 \\
  \hline
  U-1000-10 & 0.6 & 75 & 0.600 & 40 & 71 & 0.602 & 295 & 64 & 0.600 & 220 & \textbf{64} & \textbf{0.601} & 270 \\
  U-1000-10 & 0.7 & 130 & 0.700 & 40 & 125 & 0.701 & 601 & 131 & 0.700 & 291 & \textbf{124} & \textbf{0.700} & 755 \\
  U-1000-10 & 0.8 & 234 & 0.801 & 38 & \textbf{230} & \textbf{0.801} & 847 & 256 & 0.800 & 524 & 234 & 0.800 & 3031 \\
  \hline
  U-1000-50 & 0.6 & 357 & 0.601 & 264 & \textbf{356} & \textbf{0.601} & 7293 & 437 & 0.600 & 846 & 362 & 0.600 & 21600 \\
  U-1000-50 & 0.7 & \textbf{480} & \textbf{0.700} & 226 & \textbf{480} & \textbf{0.700} & 3821 & 562 & 0.700 & 1358 & 483 & 0.701 & 21600 \\
  U-1000-50 & 0.8 & \textbf{628} & \textbf{0.800} & 170 & \textbf{628} & \textbf{0.800} & 2804 & 696 & 0.801 & 1165 & 630 & 0.800 & 21600 \\
  \hline
  U-1000-100 & 0.6 & 438 & 0.601 & 539 & \textbf{437} & \textbf{0.600} & 21600 & 510 & 0.601 & 803 & 441 & 0.601 & 21600 \\
  U-1000-100 & 0.7 & \textbf{560} & \textbf{0.700} & 434 & \textbf{560} & \textbf{0.700} & 4937 & 623 & 0.700 & 788 & 563 & 0.700 & 21600 \\
  U-1000-100 & 0.8 & \textbf{695} & \textbf{0.800} & 336 & \textbf{695} & \textbf{0.800} & 6646 & 745 & 0.800 & 812 & 697 & 0.800 & 21600 \\
  \hline
\end{tabular}}
  \caption{{\small Results from the proposed GC heuristics for the Uniform distributed instances with $n \leq 1000$.}} \label{tab:t3}
\end{table}}

Now, considering the results reported in Table \ref{tab:t3} involving the Uniform distributed instances, we can state, in general terms, that the effectiveness of the VNS over the algorithms that produce its starting solution, the Ward's-GC and the \emph{k}means-GC, follow the same conclusions detached for the Normal distributed instances, although with more pronounced gains on the Uniform distributed tests. This may suggest that the Ward's-GC and the \emph{k}means-GC are less effective on the Uniform distributed instances or that those instances are harder than those following the Normal distribution.

One reason for explaining the good performance of the VNS/\emph{k}means-GC on some instances can be related with the fact that the two main methodologies being used (the \emph{k}-means and the Ward's reconstruction performed in the improvement phase of the VNS) are very different from each other, when compared with the two main methodologies in the VNS/Ward's-GC. For this reason, we may admit that the initial solution produce by the \emph{k}means-GC in the VNS framework is much different from the initial solution generated by the Ward's-GC, promoting, this way, the diversification required in any metaheuristic when addressing a strongly combinatorial problem. This argument can, in fact, explain some of the winner solutions of the VNS/\emph{k}means-GC when compared to the VNS/Ward's-GC. However, when the result of the \emph{k}means-GC is too far from the best solution, the VNS/\emph{k}means-GC can hardly get competitive results, demanding a huge effort to close the gap.

Considering this observation, we may suspect that the low performance of the \emph{k}means-GC can be related with our implementation of the \emph{k}-means heuristic. To analyze this issue, we compare the solutions returned by our \emph{k}-means procedure (described in Figure \ref{Fig4}) with those returned by the \emph{k}-means algorithm of SPSS. In these tests, we used the SPSS version 24.0.0, with the maximum number of iterations set to 999, with the "Iterate and classify" option, with convergence criterion equal to 0, and using running means. These experiments are conducted on the instances N-1000-50, N-1000-100, U-1000-50 and U-1000-1000, with $R^2T=0.6$, to which the \emph{k}means-GC obtained the worst results, when compared with the associated best solutions found. Table \ref{tab:t4} shows the results, considering the number of components (\emph{k}) in the best partitions found by the \emph{k}means-GC, reported in Tables \ref{tab:t2} and \ref{tab:t3}, for the selected instances. The tests were conducted on a different machine, with an Intel Core i5-M450 with 2.40 GHz and 4 GB RAM. The experiments were also performed under Microsoft Windows 10 operating system and our algorithms were also compiled on gfortran. The execution times are reported in seconds.

\smallskip {\small
\begin{table}
  \centering
 {\small
 \begin{tabular}{lc|ccc|ccc|}
  & & \multicolumn{3}{c}{Our \emph{k}-means} & \multicolumn{3}{c|}{SPSS's \emph{k}-means}
    \\ 
  Instances & $R^2T$ & $k$ & $R^2$ & time & $k$ & $R^2$ & time \\
  \hline
  N-1000-50 & 0.6 & 435 & 0.601 & 114.12 & 435 & 0.598 & 12.72 \\
  N-1000-100 & 0.6 & 504 & 0.600 & 167.10 & 504 & 0.550 & 30.50 \\
  U-1000-50 & 0.6 & 437 & 0.600 & 167.55 & 437 & 0.599 & 12.78 \\
  U-1000-100 & 0.6 & 510 & 0.601 & 147.12 & 510 & 0.526 & 29.02 \\
  \hline
\end{tabular}}
  \caption{{\small Comparison among the results of our \emph{k}-means implementation and those obtained using the SPSS package, for given values of \emph{k}.}} \label{tab:t4}
\end{table}}

The results in Table \ref{tab:t4} indicate that the quality of our solutions is competitive with state-of-the-art implementations of the \emph{k}-means, although requiring more time to execute.

\bigskip
We further test a few larger sized instances, considering $n=5000$ with $m=3 \mbox{ and } 5$, and $n=10000$ with $m=3$. Due to memory limitations and excessive execution time, we could not run algorithms \emph{k}means-GC and VNS/\emph{k}means-GC. The results are shown in Table \ref{tab:t5}.

\smallskip {\small
\begin{table}
  \centering
 {\footnotesize
 \begin{tabular}{lc|ccc|ccc|}
  & & \multicolumn{3}{c}{Ward's-GC} & \multicolumn{3}{c|}{VNS/Ward's-GC}
    \\ 
  Instances & $R^2T$ & $c(P)$ & $R^2(P)$ & time & $c(P)$ & $R^2(P)$ & time \\
  \hline
  N-5000-3 & 0.6 & 9 & 0.613 & 1053 & \textbf{9} & \textbf{0.617} & 1115 \\
  N-5000-3 & 0.7 & \textbf{15} & \textbf{0.711} & 1041 & \textbf{15} & \textbf{0.711} & 1062 \\
  N-5000-3 & 0.8 & 27 & 0.800 & 1074 & \textbf{27} & \textbf{0.802} & 1125 \\
  \hline
  N-5000-5 & 0.6 & 31 & 0.601 & 2047 & \textbf{30} & \textbf{0.600} & 2172 \\
  N-5000-5 & 0.7 & 60 & 0.701 & 2052 & \textbf{59} & \textbf{0.701} & 2240 \\
  N-5000-5 & 0.8 & 137 & 0.800 & 2071 & \textbf{134} & \textbf{0.800} & 3997 \\
  \hline
  N-10000-3 & 0.6 & 9 & 0.607 & 8120 & \textbf{9} & \textbf{0.610} & 8295 \\
  N-10000-3 & 0.7 & 16 & 0.710 & 8055 & \textbf{15} & \textbf{0.702} & 8243 \\
  N-10000-3 & 0.8 & 30 & 0.804 & 8082 & \textbf{29} & \textbf{0.801} & 8286 \\
  \hline
  U-5000-3 & 0.6 & 6 & 0.603 & 998 & \textbf{6} & \textbf{0.608} & 1107 \\
  U-5000-3 & 0.7 & 9 & 0.704 & 992 & \textbf{8} & \textbf{0.721} & 1620 \\
  U-5000-3 & 0.8 & 16 & 0.806 & 991 & \textbf{15} & \textbf{0.809} & 1883 \\
  \hline
  U-5000-5 & 0.6 & 19 & 0.608 & 1892 & \textbf{18} & \textbf{0.612} & 2460 \\
  U-5000-5 & 0.7 & 34 & 0.701 & 1894 & \textbf{30} & \textbf{0.704} & 2692 \\
  U-5000-5 & 0.8 & 79 & 0.800 & 1893 & \textbf{71} & \textbf{0.801} & 3992 \\
  \hline
  U-10000-3 & 0.6 & 6 & 0.604 & 7847 & \textbf{6} & \textbf{0.610} & 10342 \\
  U-10000-3 & 0.7 & 9 & 0.719 & 7888 & \textbf{8} & \textbf{0.713} & 13323 \\
  U-10000-3 & 0.8 & 16 & 0.801 & 7939 & \textbf{15} & \textbf{0.802} & 11039 \\
  \hline
\end{tabular}}
  \caption{{\small Results from the Ward's based GC heuristics for the Normal and the Uniform distributed instances with $n = 5000 \mbox{ and } 10000$.}} \label{tab:t5}
\end{table}}

The results reported in Table \ref{tab:t5} confirm most observations previously stressed involving the Ward's based GC algorithms, namely that the gains of the VNS over the Ward's-GC are more pronounced when \emph{n} and \emph{m} increase, and also more prominent on the Uniform distributed instances. In addition, when considering the execution times of the VNS/Ward's-GC, the largest share is due to the Ward's-GC execution. Still, it would be interesting to observe the results of the VNS/\emph{k}means-GC over these larger sized instances, considering the good performance of that algorithm on the results reported in Tables \ref{tab:t2} and \ref{tab:t3}, particularly on the instances involving a small number of attributes ($m \leq 5$), as in the tests comprising the larger sized class.


\bigskip
The main parameter in all the tests here conducted is the threshold $R^2T$ ratio. Two additional parameters involved in the execution of the VNS algorithms are the overall upper time limit and the $r_{\footnotesize\mbox{max}}$ that acts in the termination criteria. The first one was set to 21600 seconds, while the second one was set to 50, in all the tests. These parameters were not a matter of discussion in the present paper, and were kept fixed in all tests here performed.


\subsection{Tests on the real-world instances} \label{subsec:T-RW}

The computational tests conducted on real-world examples involve a microarray gene expression dataset and a datastream of Exchange Traded Funds (ETFs) daily prices from iShares. The upper time limit was removed when running the VNS based algorithms, while $r_{\footnotesize\mbox{max}}$ was kept unchanged, being set to 50 in all these tests.

\bigskip
The microarray example consist of a gene expression dataset from the \emph{Sacharomyces cerevisiae} considered in \cite{Thaetal2006}, for comparing gene clustering methods in microarray analysis. The dataset was used to explore how gene expression is remodeled in response to changes in extracellular environment, which includes changes in temperature, oxidation, nutrients, pH, and osmolarity, as described in \cite{Cauetal2001}. In the present paper, we do not provide any discussion on microarray analysis, but simply use the dataset to test our algorithms.

The example involves expression profiles of 1744 genes (elements) over 45 samples (attributes). The negative values in the original data indicate fold repression and the positive values indicate fold activation. The tests have been conducted over standardized values. Table \ref{tab:t6} reports the best solutions' information obtained using the four GC algorithms proposed in Section \ref{sec:Tests}.

\smallskip {\small
\begin{table}
  \centering
 {\footnotesize
 \begin{tabular}{c|ccc|ccc|ccc|ccc|}
  & \multicolumn{3}{c}{Ward's-GC} & \multicolumn{3}{c|}{VNS/Ward's-GC} & \multicolumn{3}{|c}{\emph{k}means-GC} & \multicolumn{3}{c|}{VNS/\emph{k}means-GC}
    \\ 
  $R^2T$ & $c(P)$ & $R^2(P)$ & time & $c(P)$ & $R^2(P)$ & time & $c(P)$ & $R^2(P)$ & time & $c(P)$ & $R^2(P)$ & time \\
  \hline
  0.6 & 79 & 0.600 & 1322 & 77 & 0.601 & 2479 & 68 & 0.601 & 1296 & \textbf{67} & \textbf{0.601} & 1679 \\
  0.7 & 200 & 0.700 & 1325 & \textbf{198} & \textbf{0.700} & 5823 & 212 & 0.700 & 1417 & \textbf{198} & \textbf{0.700} & 10440 \\
  0.8 & \textbf{433} & \textbf{0.800} & 1266 & \textbf{433} & \textbf{0.800} & 10738 & 486 & 0.800 & 3239 & 446 & 0.800 & 31536 \\
  \hline
\end{tabular}}
  \caption{{\small Results from the proposed GC heuristics when applied to the microarray gene expression dataset.}} \label{tab:t6}
\end{table}}

Considering the results in Table \ref{tab:t6}, the VNS/\emph{k}means-GC was more effective when the threshold is low ($R^2T=0.6$), while the VNS/Ward's-GC performed better when $R^2T=0.8$. In addition, the execution times of the VNS/\emph{k}means-GC grow much faster than for the VNS/Ward's-GC, along with the $R2T$ increase.

Table \ref{tab:t7} shows each attribute individual $R^2$ ratio, concerning the partition found by the VNS/Ward's-GC algorithm with $R^2T=0.8$, which separates the genes into 433 components.

\smallskip {\small
\begin{table}
  \centering
 {\footnotesize
 \begin{tabular}{cllc|cllc}
 $j$ & \multicolumn{2}{c}{Samples} & $R^2_j$ & $j$ & \multicolumn{2}{c}{Samples} & $R^2_j$ \\
  \hline
  1 & Heat & 0' (A) & 0.812 & 24 & msn2/4 acid & 0' & 0.724 \\
  2 & Heat & 0' (B) & 0.769 & 25 & msn2/4 acid & 10' & 0.738 \\
  3 & Heat & 15' & 0.745 & 26 & msn2/4 acid & 20' & 0.731 \\
  4 & Heat & 30' & 0.779 & 27 & Peroxide & 0' (A) & 0.788 \\
  5 & Heat & 45' & 0.813 & 28 & Peroxide & 0' (B) & 0.774 \\
  6 & Heat & 60' & 0.776 & 29 & Peroxide & 10' & 0.741 \\
  7 & Heat & 120' & 0.784 & 30 & Peroxide & 20' & 0.818 \\
  8 & Acid & 0' (A) & 0.863 & 31 & Peroxide & 40' & 0.870 \\
  9 & Acid & 0' (B) & 0.830 & 32 & Peroxide & 60' & 0.824 \\
  10 & Acid & 10' & 0.757 & 33 & Peroxide & 120' & 0.704 \\
  11 & Acid & 20' & 0.831 & 34 & NaCl & 0' & 0.849 \\
  12 & Acid & 40' & 0.786 & 35 & NaCl & 15' (A) & 0.785 \\
  13 & Acid & 60' & 0.831 & 36 & NaCl & 30' (B) & 0.847 \\
  14 & Acid & 80' & 0.820 & 37 & NaCl & 45' & 0.888 \\
  15 & Acid & 100' & 0.815 & 38 & NaCl & 60' & 0.809 \\
  16 & Alkali & 0' (A) & 0.863 & 39 & NaCl & 120' & 0.802 \\
  17 & Alkali & 0' (B) & 0.830 & 40 & Sorbitol & 0' & 0.849 \\
  18 & Alkali & 10' & 0.766 & 41 & Sorbitol & 15' & 0.812 \\
  19 & Alkali & 20' & 0.760 & 42 & Sorbitol & 30' & 0.870 \\
  20 & Alkali & 40' & 0.764 & 43 & Sorbitol & 45' & 0.833 \\
  21 & Alkali & 60' & 0.734 & 44 & Sorbitol & 90' & 0.828 \\
  22 & Alkali & 80' & 0.782 & 45 & Sorbitol & 120' & 0.846 \\
  23 & Alkali & 100' & 0.773 & & & & \\
  \hline
\end{tabular}}
  \caption{{\small $R^2_j$ ratios for each attribute $j=1, \ldots, 45$, considering the partition found by the VNS/Ward's-GC algorithm with $R^2T=0.8$.}} \label{tab:t7}
\end{table}}

The values reported in Table \ref{tab:t7}, show that for most samples and for this particular common partition, the proportion of the total variance captured within the clusters is below 0.2 (those having an individual $R^2$ ratio higher than 0.8), namely for the "Acid", the "NaCl" and the "Sorbitol" samples. However, there are still some samples to which the proportion of the total variance within the clusters is higher than 0.25 (those with $R^2_j \leq 0.75$), namely for "Heat 15'", "Alkali 60'", Peroxide 10'", Peroxide 120'" and for all the "msn2/4 acid" samples.

\begin{figure}
\begin{center}
  \includegraphics[scale=0.37]{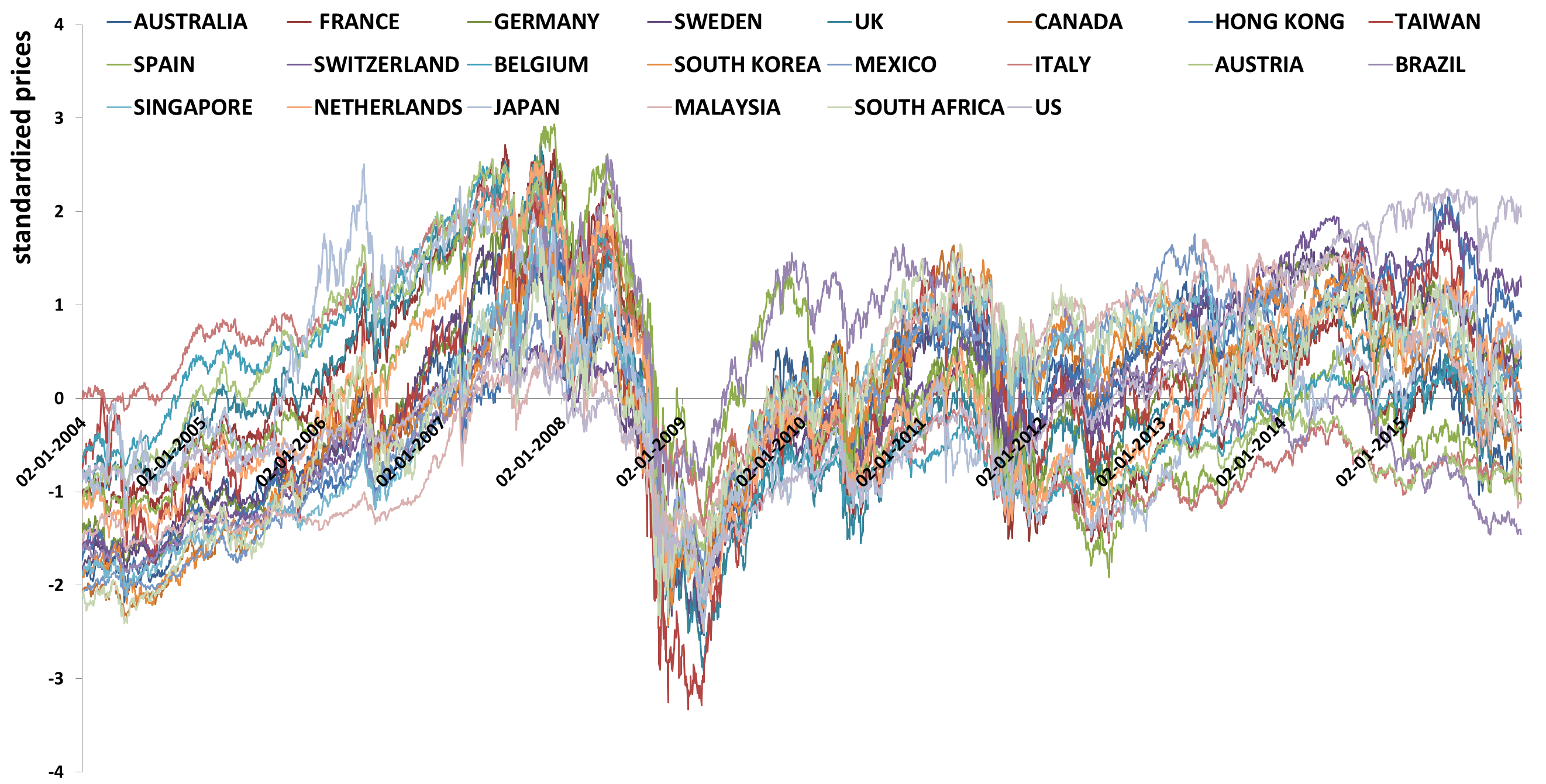}
  \caption{\footnotesize Standardized daily prices of the 22 countries' ETFs iShares in the study.}\label{Fig7}
\end{center}
\end{figure}

\bigskip
The ETFs daily prices involve 22 countries' iShares funds, ranging from 2004 to 2015, used in \cite{Nevesetal2017}. ETFs are open-ended investment funds, traded in a stock market. Its main focus is to obtain a performance related to a benchmark. It is one of the fastest growing investment products in the world. An iShare fund tracks a bond or stock market index, being the largest issuer of ETFs in most countries with dynamic economies. As described in \cite{Nevesetal2017}, the dataset was collected from the Thompson Financial Datastream, comprising the daily prices of 22 countries' ETFs from iShares, ranging from January 2, 2004 to December 31, 2015, representing 3028 days in the entire sequence. All values have been standardized. Figure \ref{Fig7} show the standardized prices along the entire time stream.

\begin{figure}
\begin{center}
  \includegraphics[scale=0.56]{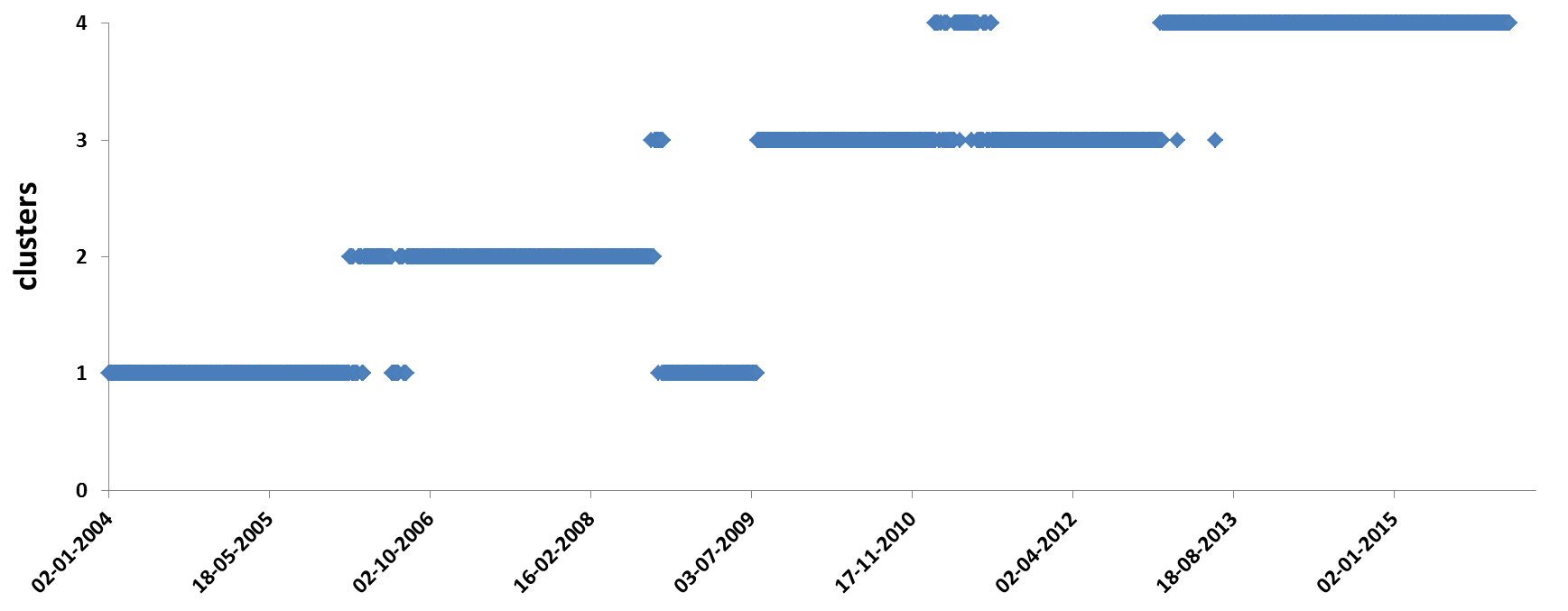}
  \caption{\footnotesize Representation of the 4 clusters in the best solution found by the VNS/\emph{k}means-GC algorithm.}\label{Fig8}
\end{center}
\end{figure}

Considering this dataset, we used the GC problem to find a partition of the trading days into clusters, such that the associated $R^2$ ratio is at least the threshold $R^2T=0.7$. The best solution was obtained by the VNS/\emph{k}means-GC algorithm. It involves 4 components, grouping the 3028 days into 4 groups of similar vectors of standardized prices. The associated $R^2$ ratio is equal to 0.727. The algorithm required 24301 seconds to reach the goal. Most of the execution time was consumed by the \emph{k}means-GC, taking 23974 seconds. Actually, the \emph{k}means-GC almost found the same solution, also involving 4 components and with $R^2=0.726$. The best solution is shown in Figure \ref{Fig8}, with the components/clusters represented in the vertical axis and the elements (trading days) represented in the horizontal axis.

If we use, instead, the VNS/Ward's-GC, the best solution also involves 4 components, but with a $R^2$ ratio equal to 0.708. In this case, the algorithm requires just 2572 seconds. However, the solution returned in its starting phase, handled by the Ward's-GC, involves 5 components, with $R^2=0.767$ and taking 1430 seconds.

Comparing Figures \ref{Fig7} and \ref{Fig8}, we can observe that the partition of the trading days produces an interesting separation of the entire time stream into long and continuous periods, separating and isolating the 2008 hard financial crisis, represented by the small stream in the first cluster, ranging from October 2, 2008 to July 22, 2009. In effect, the ETFs daily prices in that period returned back to those of 2005. In addition, we can also detach pre-crisis and post-crisis periods, including the more recent tendency to a recovery trajectory.

As observed in the Introduction section, the $R^2$ ratio provides a global indicator of the total variance captured by the partition, integrating in a single value the particular variances of each attribute on the same common partition. Table \ref{tab:t8} presents all those particular variances, represented by each $R^2_j$ for $j=1, \ldots, 22$, as described in expression \ref{eqn:c7}.

\smallskip {\small
\begin{table}
  \centering
 {\footnotesize
 \begin{tabular}{clc|clc|clc}
 $j$ & Country & $R^2_j$ & $j$ & Country & $R^2_j$ & $j$ & Country & $R^2_j$ \\
  \hline
  1 & Australia & 0.670 & 9 & Spain & 0.603 & 17 & Singapore & 0.766 \\
  2 & France & 0.740 & 10 & Switzerland & 0.881 & 18 & Netherlands & 0.748 \\
  3 & Germany & 0.757 & 11 & Belgium & 0.709 & 19 & Japan & 0.687 \\
  4 & Sweden & 0.777 & 12 & South Korea & 0.747 & 20 & Malaysia & 0.743 \\
  5 & UK & 0.655 & 13 & Mexico & 0.805 & 21 & South Africa & 0.749 \\
  6 & Canada & 0.721 & 14 & Italy & 0.791 & 22 & US & 0.754 \\
  7 & Hong Kong & 0.787 & 15 & Austria & 0.764 & & & \\
  8 & Taiwan & 0.579 & 16 & Brazil & 0.560 & & & \\
  \hline
\end{tabular}}
  \caption{{\small $R^2_j$ ratios for each attribute (country) $j=1, \ldots, 22$ in the ETFs iShares study.}} \label{tab:t8}
\end{table}}

Looking to the values reported in Table \ref{tab:t8}, we can state that most $R^2_j$ ratios are above the threshold, indicating a number of countries whose variance captured within the clusters is below 30\% (those with $R^2_j \geq 0.7$). However, and for the same partition, there are two countries that still retain almost 50\% of the entire variance within the clusters, namely Taiwan and Brazil. For these countries, the proposed common partition does not produces an adequate separation of the daily prices. That is, if the partition would have been handled individually for each country, the interval that characterizes the hard crisis period would probably be much different for these two countries.

\section{Conclusions} \label{sec:Conclusions}

This paper introduces a new methodology within non-hierarchical clustering, named Goal Clustering (GC). Instead of fixing the number of clusters/components (\emph{k}) as in the usual \emph{k}-means method, the GC problem sets in advance a given $R^2$ ratio threshold (denoted by $R^2T$) and finds a partition $P$ with the minimum number of components (\emph{k}), satisfying the condition $R^2(P) \geq R^2T$. This new methodology avoids a known drawback of the \emph{k}-means, involving the difficulty to set parameter \emph{k}. Instead, we start defining a lower limit threshold for the total variance kept between the components (parameter $R^2T$) and let the problem determine a partition having the smallest number of components, satisfying the threshold.

We have proposed two VNS based heuristics, considering two different procedures to find the initial solution of the VNS.

The computational tests conducted involved two kinds of instances: i) randomly generated and ii) real applied problems. The randomly generate instances followed two types of distributions: standard Normal ($N(0,1)$) and Uniform ($U(-1,1)$); and considering two classes for each randomly generated type: small/medium and large sized instances. In the small/medium type, the number of elements range in the interval $100 \leq n \leq 1000$ and up to 100 attributes ($m \leq 100$), while in the larger sized instances, the dimensions are: $n = 5000 \mbox{ and } 10000$ and $m = 3 \mbox{ and } 5$. The real-world problems involved a microarray gene expression dataset from the \emph{Sacharomyces cerevisiae}; and a datastream of ETFs daily prices from iShares.

Considering the randomly generated instances, the results indicate that the VNS is more effective over the solution generated by the \emph{k}means-GC than over the Ward's-GC, especially for the larger sized instances up to $n=1000$. A reason for this can be related with the weak performance of the \emph{k}means-GC. On the other hand, the VNS/\emph{k}means-GC found solutions with a smaller number of components on the instances with fewer attributes; while the VNS/Ward's-GC performed better on the instances with a larger number of attributes. However, in most cases, the execution time of the VNS/\emph{k}means-GC is much more demanding, especially when the number of elements (\emph{n}) increase. These observations are common to both Normal and Uniform distributed instances, although with slightly more pronounced gains on the Uniform distributed tests. This may suggest that the Ward's-GC and the \emph{k}means-GC are less effective on the Uniform distributed instances or that those instances are harder than those based on the Normal distribution.
On the larger sized instances, with $n \geq 5000$, we could only handle algorithm VNS/Ward's-GC due to memory limitations and excessive execution time of the VNS/\emph{k}means-GC. However, it would be interesting to observe the results of the VNS/\emph{k}means-GC over these larger sized instances, considering the good performance of that algorithm on the results reported in Tables \ref{tab:t2} and \ref{tab:t3}, particularly on the instances involving a small number of attributes ($m \leq 5$).

Concerning the real applied problems, we have been able to solve a microarray dataset involving 1744 elements (genes) and 45 attributes (samples) and a datastream of ETFs daily prices from iShares with 3028 elements (days) and 22 attributes (countries). The results were obtained in less than 9 hours for the microarray dataset and in almost 7 hours for the ETFs example, using the VNS/\emph{k}means-GC algorithm. If using the VNS/Ward's-GC, approximate solutions were obtained in less than 3 hours when addressing the microarray example; and less than 1 hour when considering the ETFs datastream. In most cases, the solutions produced by the two algorithms were close to each other, with some exceptions.

Considering the results reported, we detach the need to improve the \emph{k}-means based algorithms proposed for the GC. This can involve a more effective heuristic for the \emph{k}-means, including a revision on the \emph{p}-median procedure that generates its starting solution, whose memory space is too demanding. Nevertheless, we should stress the competitive performance of our \emph{k}-means implementation when compared with state-of-the-art implementations in the literature, namely when comparing with the \emph{k}-means procedure in the SPSS package.

Other aspects to improve include the execution times of the algorithms. To this purpose, we can obtain faster execution times if running the tests on a Linux environment. In addition, we can also discuss changes on the $r_{\footnotesize\mbox{max}}$ parameter that controls the termination criteria of the VNS.

Another aspect to explore in the future is a different GC version that sets individual threshold $R_j^2$ ratios to each attribute $j$, for $j=1, \ldots, m$. In this case, we can set a common threshold value imposed to all the attributes; or set individual threshold values, distinguishing the particular relevance of each attribute.


\section*{Acknowledgements}
This work was partially supported by the Portuguese National Funding: Fundação para a Ciência e a Tecnologia - FCT (project UID/MAT/04561/2013). We thank Professors Maria Elisabete Neves and Carla Fernandes for making available the ETFs daily prices dataset. Thanks are also due to Professor Richard A. Young for the microarray dataset.

\section*{Appendix A}

Proposition \ref{Prop1}:
\emph{Given a partition $P'$ of $V$ and its neighborhood $N(P')$. Let $P'' \in N(P')$, with $A$ and $B$ the two merged components from $P'$. Then, $R^2(P') - R^2(P'') = \frac{1}{SST} \cdot \frac{|A|.|B|}{|A|+|B|} \sum_{j=1}^m \left( \overline{x}_A^j  -  \overline{x}_B^j  \right)^2$.}

\bigskip
{\bf Proof:}

Considering expression (\ref{eqn:c1}), then $R^2(P') - R^2(P'') = \frac{1}{SST} \left( SSB(P') - SSB(P'') \right)$. In addition, we assume that $P''$ is obtained by merging components $A$ and $B$ from $P'$ (with $A, B \neq \emptyset$), while all other components are common to $P'$ and $P''$. So, for calculating $SSB(P') - SSB(P'')$ we can eliminate the terms in $SSB(P')$ and $SSB(P'')$ involving all components except $A$, $B$ and $A \cup B$, hence

\bigskip
$R^2(P') - R^2(P'') = \frac{1}{SST} \left( SSB(P') - SSB(P'') \right) =$

\begin{equation}\nonumber
= \frac{1}{SST} \left[ \sum_{j=1}^m |A| \cdot \left( \overline{x}_A^j - \overline{x}^j \right) ^2  +  \sum_{j=1}^m |B| \cdot \left( \overline{x}_B^j - \overline{x}^j \right) ^2  - \sum_{j=1}^m |A \cup B| \cdot \left( \overline{x}_{A \cup B}^j - \overline{x}^j \right) ^2 \right] =
\end{equation}

\begin{flushleft}
$= \displaystyle \frac{1}{SST} \cdot \sum_{j=1}^m \left[ |A| \cdot \left( \left(\overline{x}_A^j \right)^2 - 2 \overline{x}_A^j \overline{x}^j + \left(\overline{x}^j\right)^2 \right)  +  |B| \cdot \left( \left(\overline{x}_B^j \right)^2 - 2 \overline{x}_B^j \overline{x}^j + \left(\overline{x}^j\right)^2 \right)  \right] -$
\end{flushleft}
\begin{flushright}
$\displaystyle  - \frac{1}{SST} \cdot \sum_{j=1}^m \left[ (|A|+|B|) \cdot \left( \left(\overline{x}_{A \cup B}^j \right)^2 - 2 \overline{x}_{A \cup B}^j \overline{x}^j + \left(\overline{x}^j\right)^2 \right)  \right] =$
\end{flushright}

\begin{flushleft}
$= \displaystyle \frac{1}{SST} \cdot \sum_{j=1}^m \left[ |A| \cdot \left(\overline{x}_A^j \right)^2  +  |B| \cdot \left(\overline{x}_B^j \right)^2  -  (|A|+|B|) \cdot \left(\overline{x}_{A \cup B}^j \right)^2 \right] +$
\end{flushleft}
\begin{flushright}
$\displaystyle + \frac{1}{SST} \cdot \sum_{j=1}^m \left[ - 2 \overline{x}^j \cdot \left(|A| \overline{x}_A^j + |B| \overline{x}_B^j - (|A|+|B|) \overline{x}_{A \cup B}^j \right)  \right] =$
\end{flushright}

\begin{equation}\label{eqn:Ac1}
= \frac{1}{SST} \cdot \sum_{j=1}^m \left[ |A| \cdot \left(\overline{x}_A^j \right)^2  +  |B| \cdot \left(\overline{x}_B^j \right)^2  -  (|A|+|B|) \cdot \left(\overline{x}_{A \cup B}^j \right)^2  \right]
\end{equation}

\bigskip
As $ \displaystyle \overline{x}_{A \cup B}^j = \frac{\sum_{c \in A \cup B} x_{cj}}{|A \cup B|} = \frac{\sum_{a \in A} x_{aj} + \sum_{b \in B} x_{bj}}{|A| + |B|}$, equality (\ref{eqn:Ac1}) can be rewritten as
\bigskip

\begin{equation}\nonumber
\frac{1}{SST} \cdot \sum_{j=1}^m \left[ |A| \cdot \left(\overline{x}_A^j \right)^2  +  |B| \cdot \left(\overline{x}_B^j \right)^2  -  \frac{\left(\sum_{a \in A} x_{aj} + \sum_{b \in B} x_{bj} \right)^2}{|A|+|B|}  \right] =
\end{equation}

\begin{flushleft}
$= \displaystyle \frac{1}{SST} \cdot \sum_{j=1}^m \left[ |A| \cdot \left(\overline{x}_A^j \right)^2  +  |B| \cdot \left(\overline{x}_B^j \right)^2  -  \frac{1}{|A|+|B|} \cdot \left(\sum_{a \in A} x_{aj} \right)^2  \right] +$
\end{flushleft}
\begin{flushright}
$\displaystyle + \frac{1}{SST} \cdot \sum_{j=1}^m \left[ - 2 \cdot \frac{1}{|A|+|B|} \cdot \sum_{a \in A} x_{aj} \cdot \sum_{b \in B} x_{bj}  -  \frac{1}{|A|+|B|} \cdot \left(\sum_{b \in B} x_{bj} \right)^2  \right] =$
\end{flushright}

\begin{flushleft}
$= \displaystyle \frac{1}{SST} \cdot \sum_{j=1}^m \left[ |A| \cdot \left(\overline{x}_A^j \right)^2  +  |B| \cdot \left(\overline{x}_B^j \right)^2  -  \frac{|A|^2}{|A|+|B|} \cdot \left(\overline{x}_A^j \right)^2 \right] +$
\end{flushleft}
\begin{flushright}
$\displaystyle + \frac{1}{SST} \cdot \sum_{j=1}^m \left[ - 2 \cdot \frac{|A|.|B|}{|A|+|B|} \cdot \overline{x}_A^j \cdot \overline{x}_B^j  -  \frac{|B|^2}{|A|+|B|} \cdot \left(\overline{x}_B^j \right)^2  \right] =$
\end{flushright}

\begin{equation}\nonumber
= \frac{1}{SST} \cdot \sum_{j=1}^m \left[ \left(|A|- \frac{|A|^2}{|A|+|B|} \right) \cdot \left(\overline{x}_A^j \right)^2  -  2 \cdot \frac{|A|.|B|}{|A|+|B|} \cdot \overline{x}_A^j \cdot \overline{x}_B^j  +  \left(|B|- \frac{|B|^2}{|A|+|B|} \right) \cdot \left(\overline{x}_B^j \right)^2  \right] =
\end{equation}

\begin{equation}\nonumber
= \frac{1}{SST} \cdot \sum_{j=1}^m \left[ \frac{|A|.|B|}{|A|+|B|} \cdot \left(\overline{x}_A^j \right)^2  -  2 \cdot \frac{|A|.|B|}{|A|+|B|} \cdot \overline{x}_A^j \cdot \overline{x}_B^j  +  \frac{|A|.|B|}{|A|+|B|} \cdot \left(\overline{x}_B^j \right)^2  \right] =
\end{equation}

\begin{equation}\nonumber
= \frac{1}{SST} \cdot \frac{|A|.|B|}{|A|+|B|} \sum_{j=1}^m \left[ \left(\overline{x}_A^j \right)^2  -  2 \cdot \overline{x}_A^j \cdot \overline{x}_B^j  +  \left(\overline{x}_B^j \right)^2  \right] =
\end{equation}

\begin{equation}\label{eqn:Ac2}
= \frac{1}{SST} \cdot \frac{|A|.|B|}{|A|+|B|} \sum_{j=1}^m \left( \overline{x}_A^j  -  \overline{x}_B^j  \right)^2
\end{equation}

Showing the intended result. $\hspace*{9cm}\square$

\end{document}